# Enhancement in Magnetic and Magnetocaloric Properties of $CoFe_2O_4$ Nanofibers at Lower Temperatures.

Salma El mouloua[1], Youness Hadouch [2,*], Salma Ayadh[1], Salma Touili[1,3], Daoud Mezzane[1,3], M'barek Amjoud[1], Said Ben Moumen[1], Alimoussa Abdelhadi[1], Abdelilah Lahmar[3], Zvonko Jaglicic[4], Zdravko Kutnjak[2], Mimoun El Marssi[3].

**1** Laboratory of Innovative Materials, Energy and Sustainable Development (IMED), Cadi-Ayyad University, Faculty of Sciences and Technology, BP 549, Marrakech, Morocco.

**2** Condensed Matter Physics Department, Jozef Stefan Institute, Jamova Cesta 39, 1000 Ljubljana, Slovenia.

**3** Laboratory of Physics of Condensed Matter (LPMC), University of Picardie Jules Verne, Scientific Pole, 33 rue Saint-Leu, 80039 Amiens Cedex 1, France.

**4** Institute of Mathematics, Physics and Mechanics & University of Ljubljana, Faculty of Civil and Geodetic Engineering, Jamova cesta 2, 1000 Ljubljana, Slovenia.

***Corresponding author:**

E-mail: hadouch.younes@gmail.com; youness.hadouch@ijs.si

ORCID: https://orcid.org/0000-0002-8087-9494

Tel: +212-6 49 97 06 74

***Highlights:***

- Uniform $CoFe_2O_4$ nanofibers (CFO NFs) have been prepared by electrospinning combined with sol–gel process.
- High saturation magnetization was achieved at low temperatures in CFO by forming NFs.
- Magnetic properties of the CFO NFs are superior to those of the corresponding nanoparticle counterparts.
- A high magnetic entropy change $|\Delta S|$ and large relative cooling power (RCP) were obtained at low temperatures.

## Abstract

This research paper investigates new and first insights into the magnetic and magnetocaloric properties of one-dimensional (1D) cobalt ferrite $CoFe_2O_4$ (CFO) nanofibers elaborated by sol-gel based electrospinning technique, particularly focusing on their behavior at low temperatures for specific applications. The calcined CFO nanofibers' microstructural, structural, magnetic, and magnetocaloric properties were explored. The nanofibers' microstructure, with an average diameter of 210 nm, was examined by scanning and transmission electron microscopies (SEM, TEM). The X-ray diffraction (XRD) of the CFO nanofibers showed a pure cubic close-packed (c.c.p) spinel crystalline structure with the Fd $\bar{3}$ m space group. The Raman spectroscopic studies further confirm the cubic inverse spinel phase. The Magnetic properties were explored as a function of temperature, ranging from 10 to 300 K, a ferromagnetic behaviour was observed with the highest saturation magnetization of 75.87 emu/g and a coercivity of 723 Oe at room temperature. The variation of the magnetic entropy was measured indirectly using the Maxwell approach with an increasing magnetic field. A maximum of $|\Delta S|$=1.71 J/.K was reached around 32 K. At 180 K, the associated adiabatic temperature change, $\Delta T_{max}$, was 0.93 K, with a large RCP value of 7.58 J/kg was measured, which is reasonably high for the corresponding nanoparticles (NPs). This work may suggest that 1D CFO nanofibers offer a promising route for the production of nanostructured magnetic materials, potentially impacting various electronic and electromagnetic device applications at low temperatures.

***Acknowledgements:***

The authors gratefully acknowledge the generous financial support of ?????

***Formatting of funding sources:***

## 1. Introduction

In the face of accelerating climate change, the need for energy-efficient and ecologically friendly refrigeration for the conventional gas compression and expansion cooling systems is greater than ever before [1]. The Magnetocaloric Effect (MCE) technology provides a viable answer with its possible application in green magnetic refrigerators that might considerably reduce our dependency on conventional cooling systems. The MCE is being used not just for room-temperature refrigeration but also for a number of other cooling applications ranging from low-temperature to room-temperature refrigeration [2]. By definition, the magnetocaloric effect describes the ability of magnetic materials to either absorb or release heat energy from or to their surroundings during magnetization and demagnetization cycles [1]. The MCE is a property of all magnetic materials that was discovered before 1917 by the French and by the Swiss physicists Weiss and Piccard, respectively [3]. When a magnetic field is applied to a sample, it changes the magnetic state as well as the structural arrangement, resulting in a change in magnetic entropy ($\Delta S$) [4]. Consequently, the MCE is defined as an isothermal change in entropy $\Delta S$ (or an adiabatic change in temperature $\Delta T$) caused by an external magnetic field, H [5]. Several kinds of magnetic materials listed in the literature are found to be showing a MCE, including perovskite manganites like $LaMnO_3$ [6], and spinel ferrites such as $MFe_2O_4$ (where M represents a divalent ions, commonly known as $Mg^{2+}$, $Co^{2+}$, $Ni^{2+}$) [7], [8]. Using nanosized spinel ferrites in magnetic refrigeration is considered attractive and magnetocaloric research has become a growing field. The reduced size of magnetic materials provides unique features related to changes in the magnetization process [9]. In addition, nanostructured magnetic materials have particular magnetic properties that differ from those of bulk (3D) materials. As result, the high surface-to-volume ratio of the nanostructured magnetic materials (0D, 1D and 2D) contributes to a strong size-dependent behavior in magnetization reversal, highlighting their distinct properties [10]. These properties like blocking behavior, nanoscale confinement, and nanomagnetism are important from both fundamental and possible technological, such as gas sensors [11], high-density data storage [12], spin-electronics [13], magnetic resonance imaging [12], and magnetically guided drug delivery systems [14]. Spinel ferrites with the general formula $MFe_2O_4$, where M represents a metal ion from the 3d transition elements such as (Zn, Co, Cu, Ni...) have garnered a lot of interest due to their magnetic properties. Spinel ferrites have 32 oxygen atoms in their unit cell, with two lattice sites available for cations distribution. M ions occupy one half of the octahedral coordination sites, while $Fe^{3+}$ ions are equally distributed among the other half of the octahedral (B) and tetrahedral (A) sites [15]. Among several ferrites, cobalt ferrite magnetic $CoFe_2O_4$ (CFO) is gaining popularity because of its high coercivity, magnetocrystalline anisotropy, mild saturation magnetization, chemical stability, wear resistance, and electrical insulation [16]. The CFO has an inverse spinel structure, with $Co^{2+}$ ions occupying octahedral and tetrahedral sites, and half of the $Fe^{3+}$ occupying tetrahedral sites while the other half is occupying the octahedral sites [17]. The majority of CFO ferrite's magnetic characteristics are governed by the microstructure of the CFO nanoparticles, which is closely related to the process of synthesis. Moreover, until yet to our knowledge, no study was devoted to investigating the CFO NFs ferrite's magnetocaloric properties at low temperatures. Several methods, including sol-gel auto combustion [7], hydrothermal method [18], and co-precipitation [19], have been utilized to produce cobalt ferrite. New approaches for the synthesis and production of one-dimensional (1D) magnetic materials (nanorods, nanowires, and nanofibers) have therefore received the attention of scientific communities for potential technical applications, particularly in nanotechnology [20]. Nanofibers (NFs) with a one-dimensional (1D) structure have attracted attention for their high-performance properties owing to their high surface-to-volume ratio, and specific surface area [21]. This unique property contributes to enhanced permeability compared to their nanoparticle counterparts of equivalent volume. In addition, due to their shape anisotropy one dimensional nanofibers have a higher ferromagnetic resonance frequency [22]. The electrospinning technique is widely known as an effective method for the continuous and cost-effective production of nanofibers, owing to its simplicity and ease of control. This process is straightforward and adaptable, making it suitable for generating polymer nanofibers from most polymer solutions. Furthermore, by introducing small amounts of polymers into non-spinnable inorganic precursors, electrospinning allows the creation of inorganic nanofibers [23].

The current study has a dual-fold major goal. Firstly, it aims to elaborate Nanofibers cobalt ferrite $CoFe_2O_4$ using the electrospinning method. Secondly, it intends to investigate their structural, morphological, magnetic, and magnetocaloric properties at low temperatures. Therefore, this work marks the first investigation of the magnetocaloric effect of the CFO NFs.

## 2. Material and methods

### 2.1. Material synthesis

Iron nitrate nonahydrate ($Fe(NO_3)_3,9H_2O$, Oxford)(≥99.0%), cobalt nitrate hexahydrate ($Co(NO_3)_2,6H_2O$, Alfa Aesar)(≥98.0%)), ethanol absolute (EtOH) [$CH_3CH_2OH$, Biosmart] (≥99.9%)), polyvinylpyrrolidone (PVP) (Alfa Aesar, M.W. 1 300 000), and Dimethylformamide (DMF) [$C_3H_7NO$, Sigma Aldrich] (≥99.8%), were used as raw materials to prepare CFO/PVP electrospinning solution.

To make the CFO solution, 2 mmol of iron nitrate nonahydrate and 1 mmol of cobalt nitrate hexahydrate were dissolved in EtOH and stirred continuously for 2 hours. The PVP polymer was dissolved in EtOH to make the PVP solution. Then the CFO/PVP electrospinning solution was made by combining the CFO and PVP solutions. The dropwise addition of the DMF solution under vigorous stirring adjusted the homogeneous solution.

Electrospinning of the viscous solution CFO/PVP was performed using a glass syringe and a metal needle. A 12 cm distance was maintained between the needle's tip and the collecting surface during electrospinning at a DC voltage of 25 kV. A 500-rpm drum rotation speed and a feeding rate of 0.4 ml/h were used. The as-spun NFs were dried at 80 °C under vacuum for 12 h before being annealed at 800 °C for 4 h in an air atmosphere. The electrospinning set-up is described in Fig. 1.

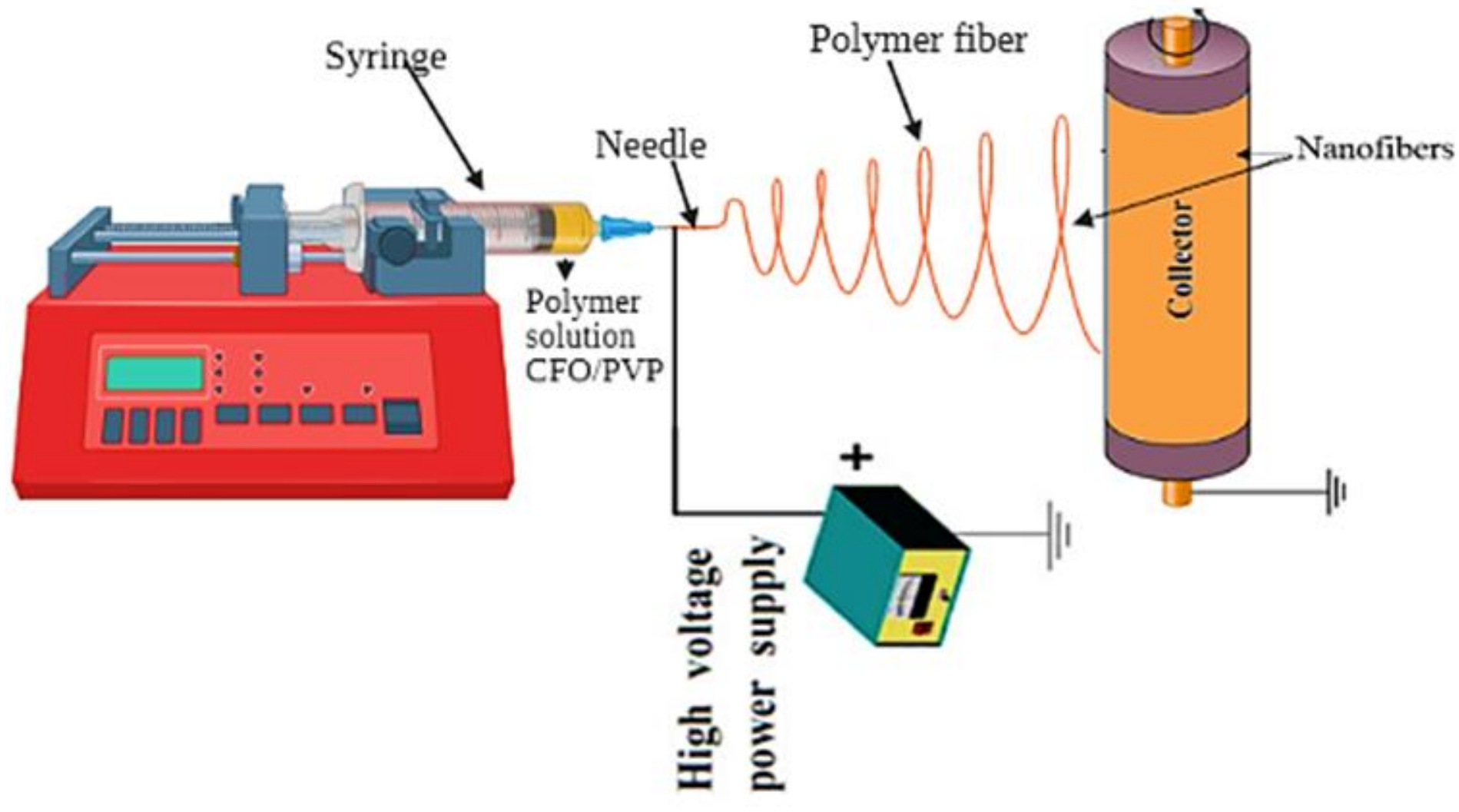

**Fig.1.** Electrospinning of CFO/PVP: setup.

### 2.2. Characterization

The thermal analysis (TGA) was performed in a Labsys Evo under an air environment at a heating rate of 5 °C/min from room temperature to 800 °C. The XRD patterns of CFO NFs were performed at room temperature on a Rigaku SmartLab SE with Cu-Kα radiation ($\lambda_{K\alpha} = 1.54056$ Å) at a scan rate of 2°/min and an angular scan range (2θ) between 20 and 80°. The structural investigations were done by fitting the XRD data using the Full-Prof software. The nanofibers microstructure was analyzed using FEI Quanta 200 scanning electron microscopy (SEM) and a JEOL – ARM200F Cold FEG high-resolution analytical transmission electron microscope (TEM) operating at 200KV. The fiber diameter distribution was estimated by ImageJ® software. The Raman spectra were recorded using a micro-Raman Renishaw spectrometer equipped with a CCD detector. The isotherm magnetization curves were measured at different temperatures 10 K–300 K, under a magnetic field variation of ±35 kOe using a Quantum Design MPMS-XL-5 SQUID magnetometer. Finally, the indirect method was used to determine the magnetocaloric temperature change based on the recorded (M-H) hysteresis as a function of temperature.

## 3. Results and discussion

### 3.1. Thermal Property of the Precursor

The result of TGA of the as-spun (uncalcined) CFO nanofibers is shown in Figure 2. Below 550 °C, the majority of the organic material (PVP), the nitrate group ($NO_3^-$) of inorganic precursors, and the other volatiles (ethanol, $H_2O$, etc.) were eliminated. The first loss was attributed to non-structural water and the decomposition of ethanol. The second loss corresponded to the decomposition of inorganic salts and the degradation of organic compounds such as PVP and DMF. The TGA analysis shows that annealing the CFO nanoparticles at 800°C provides higher thermal stability because no masse change is observed above 650°C. This temperature proved to be optimal to produce a single-phase compound with good crystallinity.

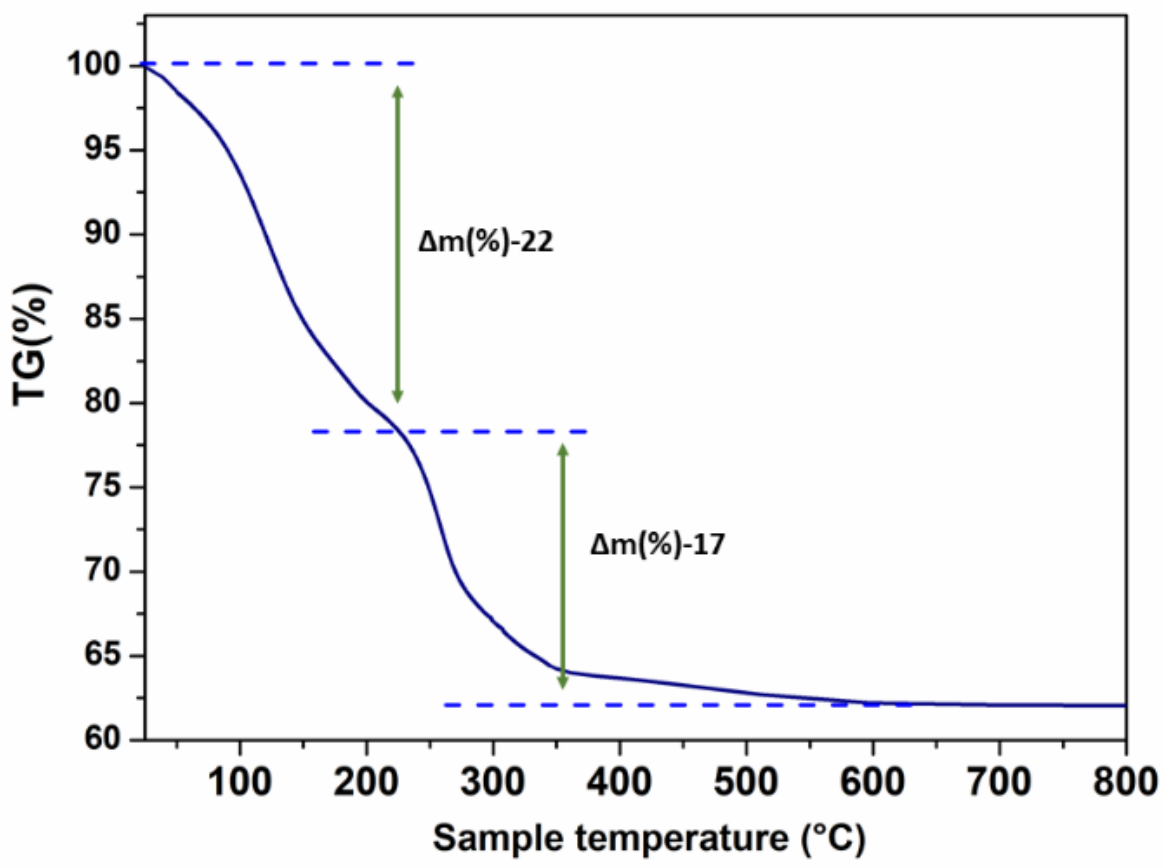

**Fig.2.** TGA curve of precursor NFs PVP/$Fe(NO)_3$, $Co(NO)_3$

**3.2. Morphological and Structural Analysis:**

SEM was used to examine the surface morphology of fibers formed after thermal annealing at 800°C for 4 h. As seen in Figs.3(a), the resultant CFO fibers maintained their continuous and uniform fibrs shape. Using ImageJ® software, an average fiber diameter of 210 nm was estimated as can be seen in Fig.3(c). The magnified region Fig.3(b) shows that CFO fiber is made up of tiny grains measuring tens of nanometers, and all of the fibers have rough features as result of the removal of PVP from the as-spun fibers during annealing process at 800 °C.

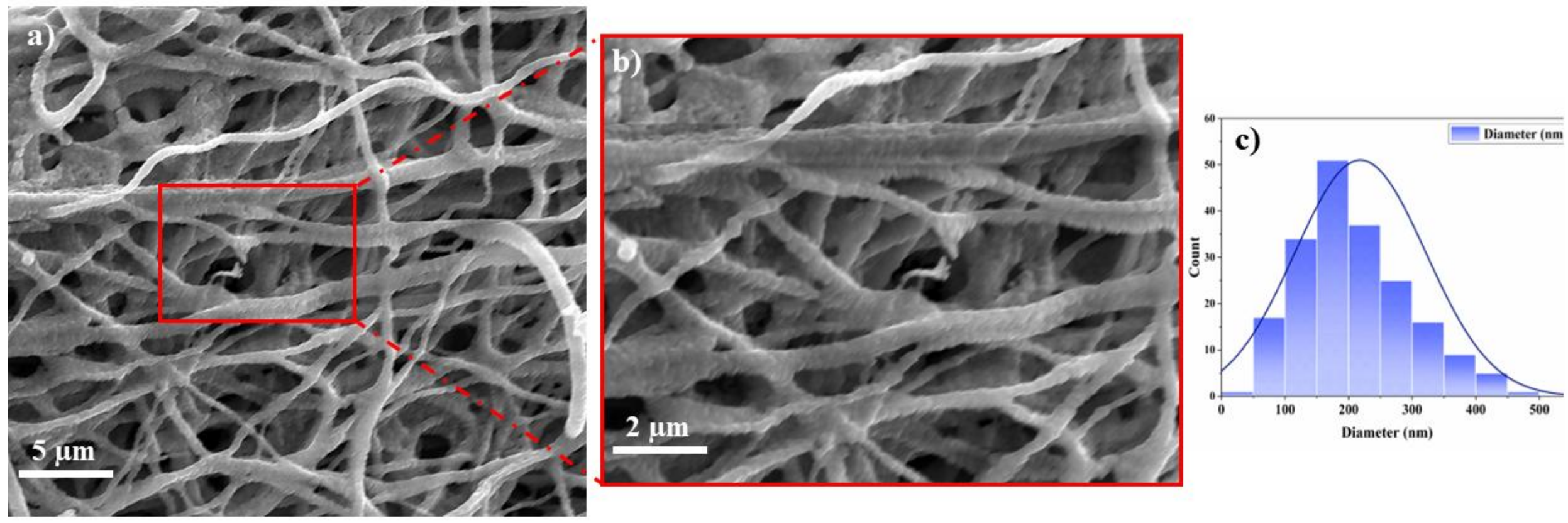

**Fig.3.** (a, b) SEM image of CFO fibers annealed at 800°C for 4 h c) fiber-diameter distribution of electrospun fibers CFO calcined at 800 ◦C/4h.

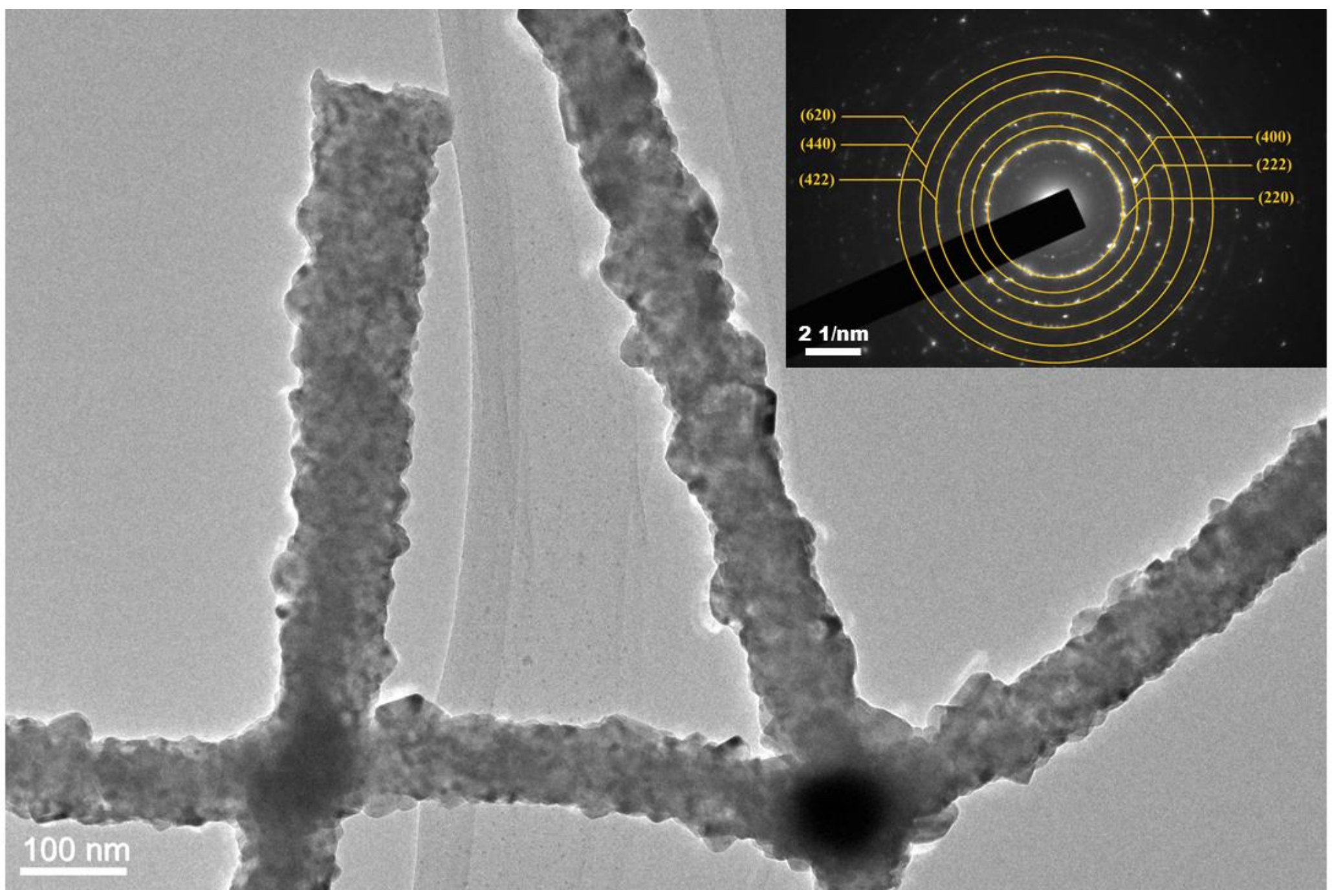

**Fig.4** TEM image of CFO NFs, inset: SAED diffraction pattern.

The CFO NFs microstructures were further observed by TEM. Fig. 4 depicts a TEM image of rough-surfaced CFO nanofibers composed of individually connected CFO nanocrystallites, the diameter of the selected CFO nanofiber is about **100** nm ???. The inset of Fig.4 displays the CFO NFs' Selected Area Electron Diffraction (SAED) pattern. Spotty ring patterns are seen in the corresponding SAED patterns of $CoFe_2O_4$ nanofibers, which are similar to the spinel ferrite structure's typical nanocrystalline nature. The SAED pattern displays six distinct diffraction rings (220), (222), (400), (422), (440), (620), which are assigned to CFO structure [19].

Fig.5(a) shows the XRD patterns of CFO fibers calcined at 800 °C/4h. The crystalline phases detected correspond to the cubic close-packed (c.c.p) structure of the spinel phase of CFO with space group Fd $\bar{3}$ m in accordance with JCPDS N°s 22-1086 [24]. The major characteristic peak that corresponds to the spinel phase CFO at 35.5° was assigned to the (311) plane. The other identified peaks (220), (222), (400), (422), (511), (440), (531), (620), (533), (622), and (444) are in agreement with the crystalline phase of the spinal cobalt ferrite system. The absence of an impurity peak indicated that the material was pure CFO fibers. By using the Full-Prof approach to fit the XRD data, more structural investigations are carried out, leading to a better match between the estimated and observed intensities with lower values of $R_p$, $R_{wp}$, and $\chi^2$ confirming the high quality of the generated samples. Fig.5(b) illustrates the Rietveld refinement profiles of CFO nanofibers and Table 1 displays the refined parameters. The goodness of the fit $\chi^2$ between the fit model and the experimental data show that the parameters obtained parameters allow to estimate a good fitting.

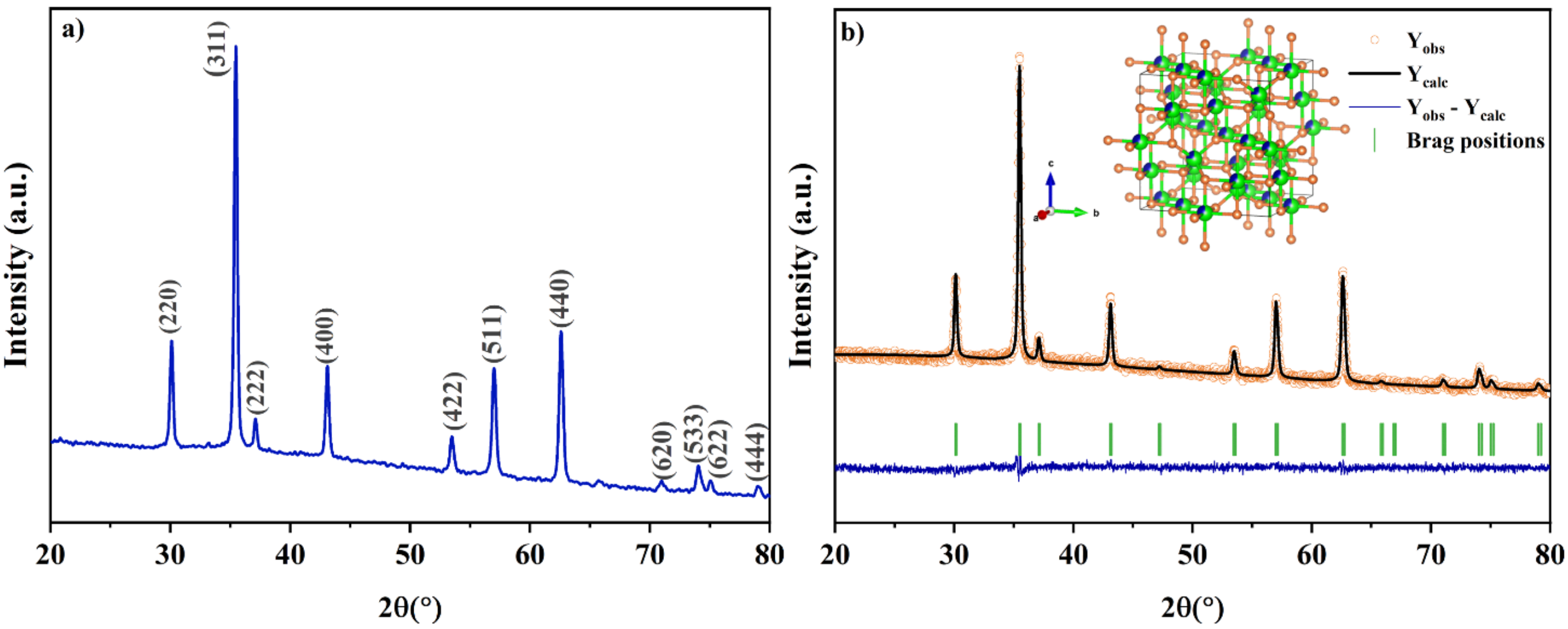

**Fig.5. a)** X-ray diffraction patterns of CFO fibers **b)** Rietveld fitted X-ray diffraction patterns of CFO nanofibers

**Table 1:** Refined structural parameters for CFO NFs at room temperature.

| Structure | Lattice parameters | | | Volume (A°)$^3$ | Reliability factors and goodness of fit | | |
|---|---|---|---|---|---|---|---|
| | a (Å) | b (Å) | c (Å) | | $\chi^2$ | $R_p$ | $R_{wp}$ |
| Fd $\bar{3}$ m | 8.3909 | 8.3909 | 8.3909 | 590.785 | 2.5601 | 0.856 | 1.08 |

To have more insight on nanocrystal structure of CFO nanofibers, Raman spectroscopy was performed over a wide spectral range from 100 to 900 cm$^{-1}$, as shown in Fig.6. For spinel ferrites (CFO) with Fd$\bar{3}$m space group, group theory analysis predicts five Raman active modes ($A_{1g}$ + $E_g$ + $3T_{2g}$), which show the presence of oxygen ion mobility and A-site and B-site within the CFO spinel structure [25]. Thus, the vibrational modes corresponding to the motion or vibration of a tetrahedral or octahedral cation are considered to be cationic occupation fingerprints, and changes in the Raman peaks as frequency shift, intensity variation, and/or band broadening are an indication for the cationic redistribution in the tetrahedral and octahedral sites [26]. It can be seen from fig.6 that the synthesized CFO spinel ferrite nanofibers show Raman modes $T_{2g}(3)$, $E_g$, $T_{2g}(2)$, $T_{2g}(1)$, $A_{1g}(2)$, and $A_{1g}(1)$ located at 210, 304, 465, 581, 615, and 686 cm$^{-1}$. The $A_{1g}(2)$ and $A_{1g}(1)$ modes are assigned frequencies of 615 and 686 cm$^{-1}$, respectively, to the Fe-O and Co-O bond stretching vibrations at tetrahedral sites under ambient temperature. The symmetric and asymmetric bending of the oxygen anions in the octahedral sublattice was represented by the modes at 304 and 581 cm$^{-1}$ [27]. The $T_{2g}(2)$ mode at 465 cm$^{-1}$ corresponds to the motion of oxygen atoms coordinated with octahedral $Fe^{3+}$ ($FeO_6$) [26], and the $T_{2g}(3)$ mode at 210 cm$^{-1}$ is associated with the translational shift of the entire $FeO_4$ tetrahedron [7], [28]. This result further supports the previously mentioned statement from the XRD and MET analyses, confirming that nanocrystals of CFO NFs were successfully achieved. Additionally, it indicates the absence of any secondary phases.

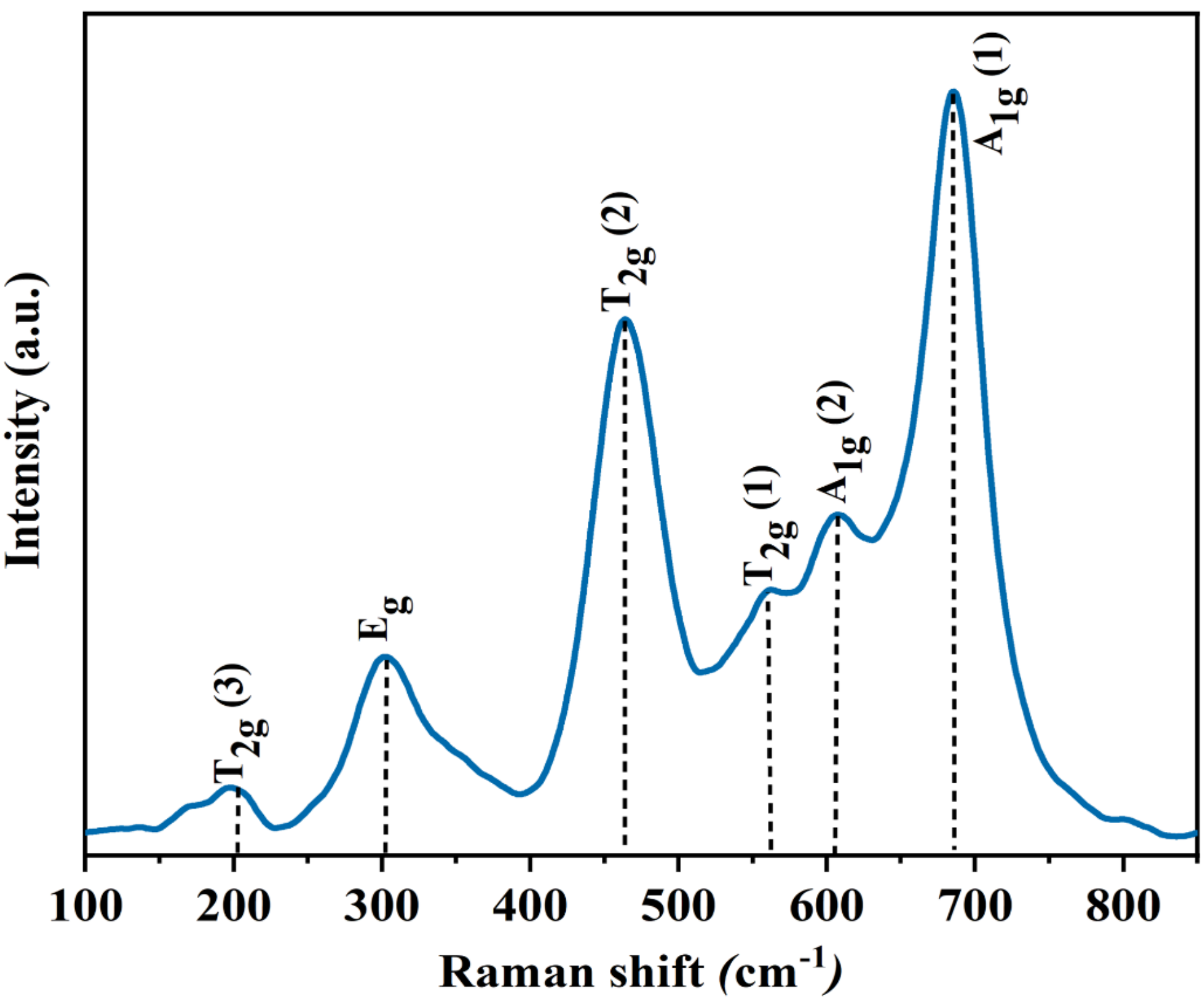

**Fig.6.** Raman spectrum of CFO nanofibers

### 3.3. Magnetic properties.

Fig.7(a) shows magnetic hysteresis (M-H) loops of the CFO nanofibers at low temperature. As we can see, a notable change was observed in the shape of the M-H hysteresis as the temperature decreased from RT to 10K. The NFs CFO seems to undergo a change from soft to hard magnetic material while decreasing temperature. The evolution of the coercivity as a function of temperature is depicted in Fig.7(c). The small coercivity at room temperature could be linked to the small size nature of these nanoparticles forming NFs [29]. However, at low temperatures, magnetic anisotropy increases and particles scatter in the direction of the anisotropic field, leading coercivity to increase [30], [31]. The large coercivity and low saturation magnetization at low temperature are consistent with a pronounced growth of magnetic anisotropy inhibiting the alignment of the moment in an applied field [32]. As can be seen in Fig.7(b, c), the saturation magnetization gradually increases with temperature from 10 to 195 K before decreasing at 200 K. The unusual behavior seen in the temperature range of 200 to 300 K is close to the irreversibility temperature ($T_{irr}$) of 195 ± 5K K observed in ZFC-FC curves as shown in fig.8. However, the anomalous magnetization near irreversibility temperature might be arise from one or more of the following effects, such as the cooperative effects of exchange springs, exchange coupling, dipolar interaction, and anisotropy, which change the energy barrier distribution [1]. It should be noted that our sample's magnetic properties are more significant than those of CFOs using other synthesis methods, as summarized in Table 2. These results might be due to fibrous systems having grains arranged and aligned in a linear chain configuration (high A/V ratio), and their magnetic dipoles will also be aligned along the fiber axis [33], [34]. Nanofibers (NFs) have different magnetic characteristics than nanoparticles (NPs) because of higher magnetocrystalline anisotropy caused by inter-particle interactions and unique geometrical shapes [35], [36]. $CoFe_2O_4$ NPs typically exhibit uniaxial single-domain behavior, whereas NFs exhibit multidomain properties with greater magnetocrystalline anisotropy [35]. The elongated structure of one-dimensional magnetic nanomaterials such as nanofibers highlights the importance of shape anisotropy, which has a substantial impact on their magnetic behavior [10]. Furthermore, the grain size and morphology of magnetic materials play an important role in determining their magnetic properties, with interactions between neighboring grain dipoles and their alignment contributing to enhanced magnetic properties [10], [34].

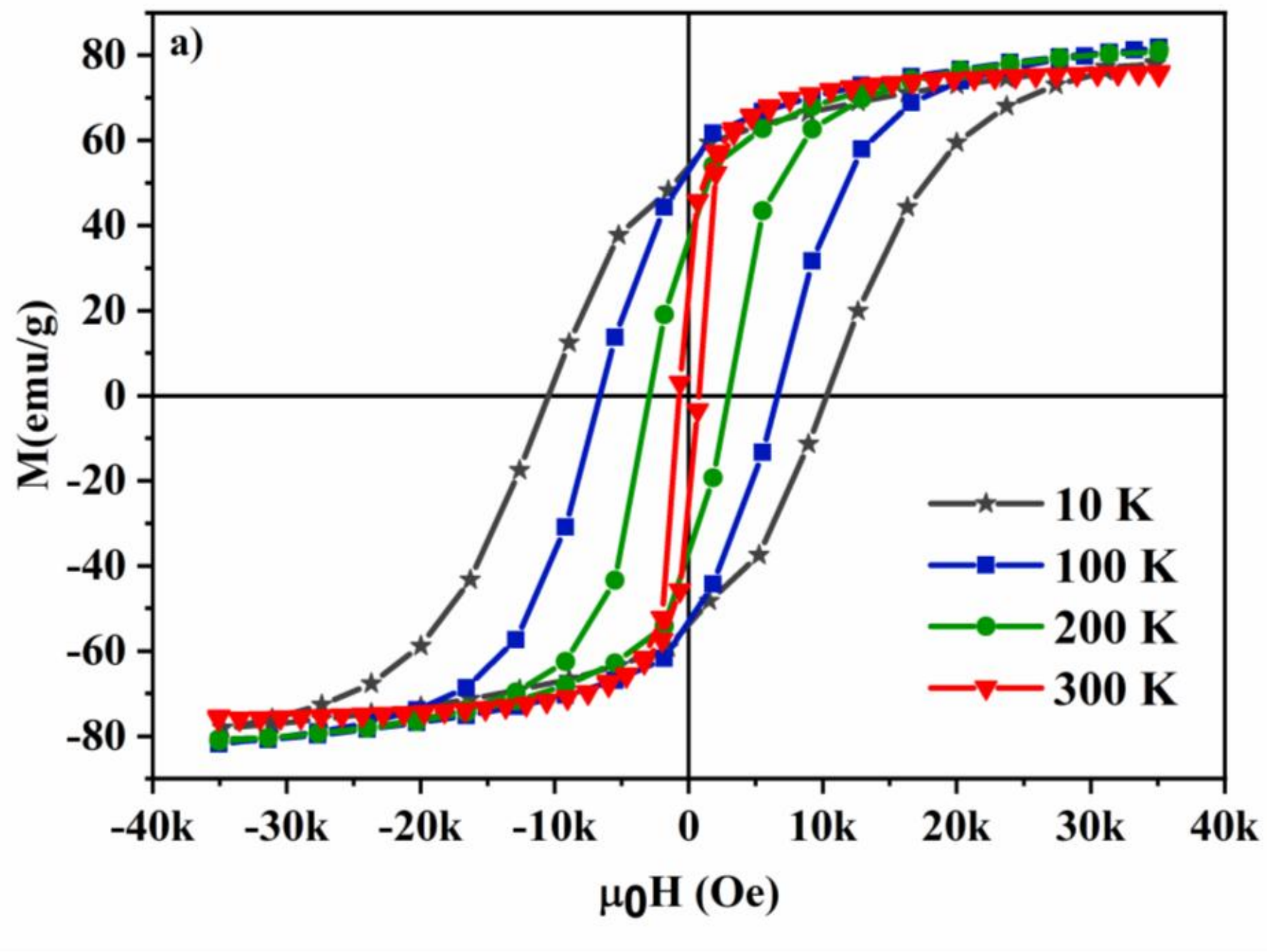

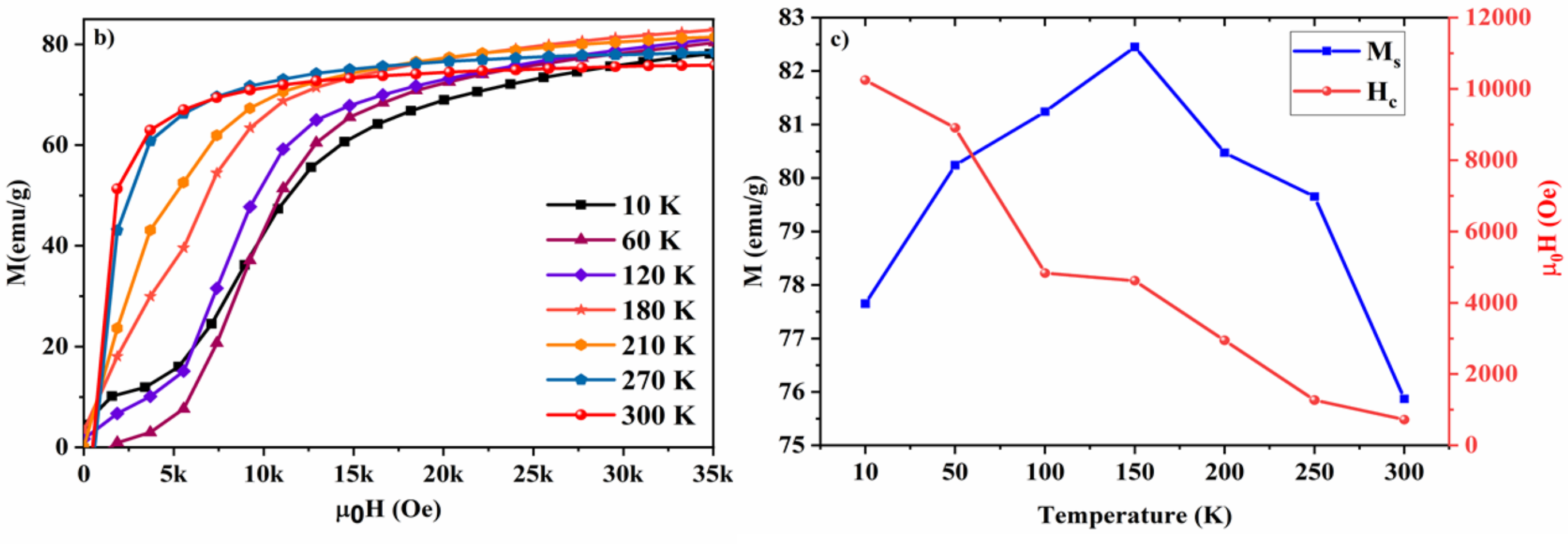

**Fig.7.** a) Hysteresis loops for the $CoFe_2O_4$ nanofibers. b) M-H plot of $CoFe_2O_4$ NFs at different temperatures. c) Temperature dependence of saturation magnetization ($M_S$) and coercivity ($H_C$) of $CoFe_2O_4$ nanofibers.

**Table 2:** Magnetic properties $M_s$ and $H_c$ at 300 K of $CoFe_2O_4$ obtained by various synthesis processes.

| Synthesis method | Particle's shape | $M_s$ (emu/g) | $H_c$ (Oe) | Magnetic field (Oe) | References |
| --- | --- | --- | --- | --- | --- |
| Electrospinning | NFs | 75.87 | 723 | 35000 | This work |
| Electrospinning | NFs | 74.54 | 723 | 20000 | This work |
| Electrospinning | NFs | 63.9 | 1098.1 | 4000 | [10] |
| Hydrothermal method | NPs | 59.02 | - | 4000 | [37] |
| Coprecipitation | NPs | 55.8 | 850 | | [38] |
| Sol-gel | NPs | 66.7 | 1163.9 | 4000 | [10] |
| Sol-gel combustion | NPs | 65.7 | 1243 | 20000 | [39] |
| Sol–gel method combined with celloulose template | Fiber | 64.97 | 1005 | 12000 | [40] |

The temperature dependence of the magnetic properties of CFO NFs was investigated by measuring the magnetic moment under zero-field-cooled (ZFC) and field-cooled (FC) conditions. The sample was cooled from room temperature to 0 K without an external magnetic field and then the magnetization (ZFC) of CFO NFs was recorded during heating up to 300 K. Considering that CFO NFs exhibit a high magnetic anisotropy, magnetic fields of 1, 5, and 10 kOe were applied during both FC and ZFC conditions. As seen in Fig. 8, ZFC magnetization increases with temperature, for all applied magnetic fields, until it reaches a critical point known as the blocking temperature $T_b$, after which it tends to decrease. This temperature was quite apparent in both the 5 and 10 kOe magnetic fields. The decrease in magnetization above the blocking temperature was related to the spin glass behavior of highly interacting particles in a magnetic system [41], this is characteristic of superparamagnetic materials [42]. The $T_b$ depends on the material, the size of the particles, and also on the presence of interparticle magnetic interactions [31]. However, FC magnetization increases slightly at all the magnetic fields corresponding to non-interacting regions [11]. The FC and ZFC curves bifurcate due to the introduction of a high magnetic field, and the magnetization in FC is higher than in ZFC conditions below $T_b$. This irreversible behavior is driven by an opposition between the energy required for structural reorientation and the energies of magnetoelasticity, form, and crystalline anisotropy[43], [44].

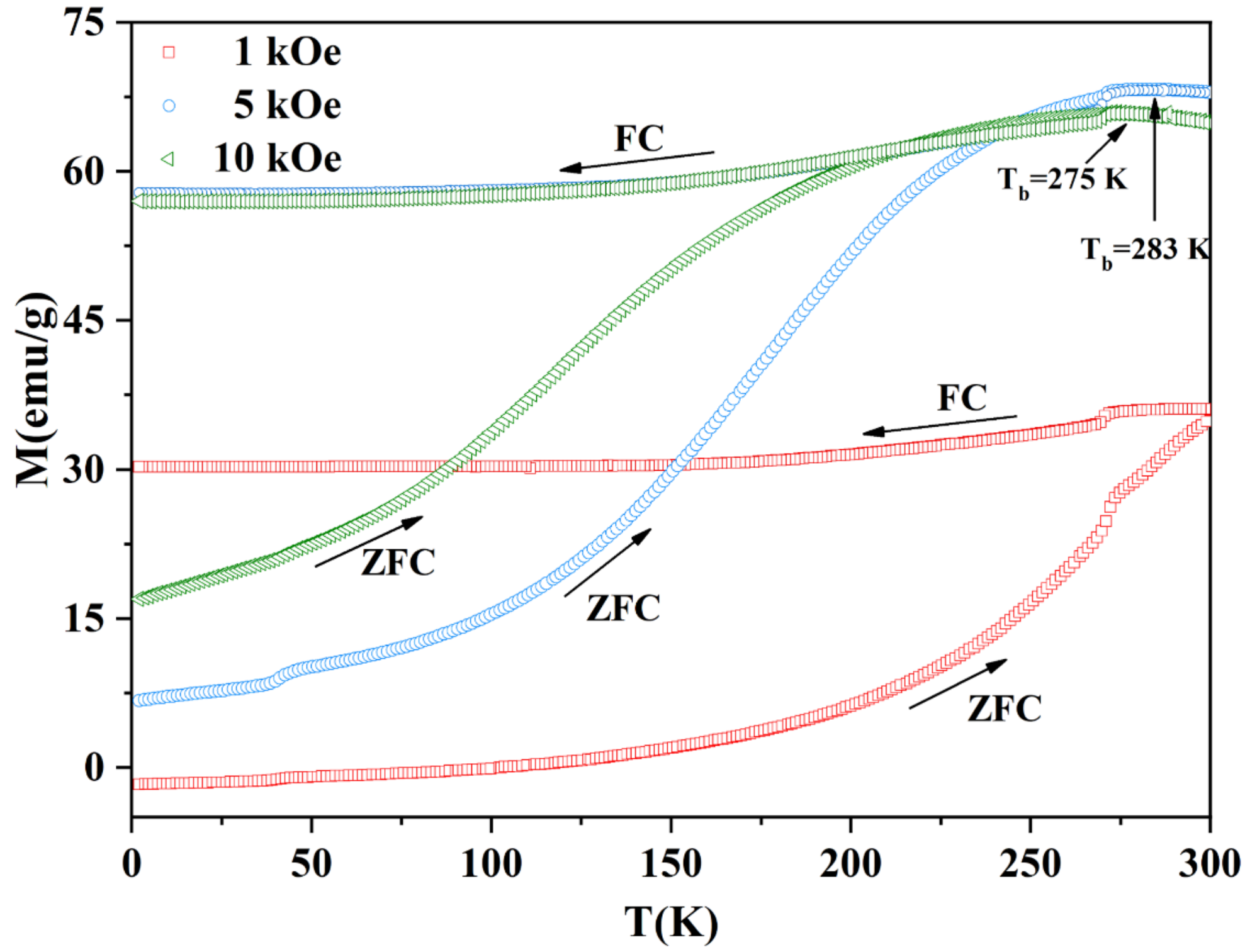

**Fig.8.** Zero field cooled and field cooled (ZFC-FC) curves of CFO NFs at H = 1 kOe; 5 kOe; and 10 kOe.

### 3.4. Magnetocaloric effect

The magnetocaloric effect can be obtained indirectly from a set of magnetic measurements as a function of temperature and magnetic field. The magnetic entropy variation ($\Delta S$) and the adiabatic temperature changes (ΔT) versus the magnetic field application can be calculated through the well-known integral which follows from the differential Maxwell thermodynamic relations Eq 1 and 2, respectively [45], expressed by the following equations:

$$\Delta S(T,H) = \int_0^H \left(\frac{\partial M}{\partial T}\right)_H \partial H. \qquad \text{Eq(1)}$$

$$\Delta T(T,H) = \int_{H_1}^{H_2} \frac{T}{C_p}\left(\frac{\partial M}{\partial T}\right)_H \partial H. \qquad \text{Eq(2).}$$

With $H_2$–$H_1$ is the change of the applied field and Cp is the specific heat capacity of the sample. The entropy change vs. temperature curve for CFO nanofibers is displayed in Fig. 9(a). The ($\Delta S$) was numerically computed using the Maxwell equation and the data sets are displayed in Fig. 7(b). Our experimental results suggest that this compound exhibits both conventional (MCE) and inverse (IMCE) types of magnetocaloric effects [46]. Magnetic entropy change attains a maximum -ΔS of −1.71 J/K.kg around 32 K. Fig.9(b) shows the magnetic adiabatic temperature change (ΔT) as a function of temperature under different applied magnetic fields. Below 180 K, ΔT exhibits similar negative characteristics as the magnetic entropy change (ΔS), with maximum values of -0.74 K. As seen in Fig. 9(b), the sign of ΔT changes to positive values regarding high values of 0.93 K around 180 K. We note that the magnetocaloric potentials; (ΔS); are negative, this behavior is known as the inverse magnetocaloric effect (IMCE). This IMCE is expected to occur in all antiferromagnetic materials [47]. As the magnetic field increases, the negative value of ΔS also rises. It is interesting to note that the sign of entropy changes from negative to positive around 120 K. Normal and inverse magnetocaloric effects are observed in regimes dominated by ferromagnetic/antiferromagnetic interactions. This kind of behavior is also observed in ball milled zinc ferrite nanoparticles at low temperature regime [48]. Similarly, inverse and normal magnetic entropy changes are reported for the $CoFe_2O_4/CoFe_2$ composite [1]. Some fluctuations are observed in ΔS that could be related to the shape change of the initial magnetization curves Fig. 7(b) at low temperatures [49]. Furthermore, fluctuations in the apparent entropy change may be influenced by factors, such as changing between various experimental conditions or starting new measurements at different temperatures. The entropy change values demonstrated in this work are significantly higher than those previously reported for various other iron oxide-based nanoparticle systems, as shown in Table 3. The highest entropy change is influenced by sample morphology, as suggested by the increased entropy change resulting from the aligned grain structure within the fibrous sample.

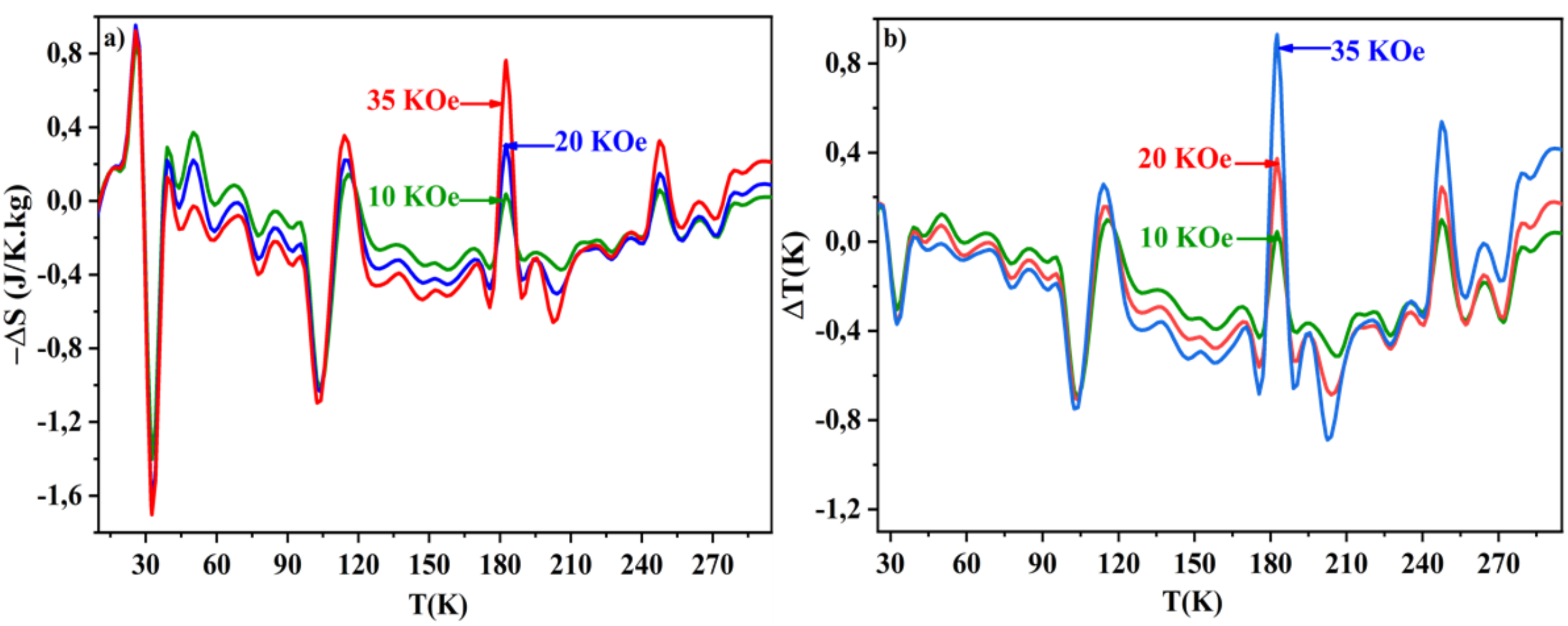

**Fig.9. a)** Magnetic entropies change ΔS **b)** The magnetic adiabatic temperature change ΔT as a function of temperature at different applied magnetic fields for CFO NFs.

**Table 3**: Magnetocaloric properties of ferromagnetic oxides

| Materials | \|ΔS\| | | \|ΔT\| | | RCP | Magnetic field (Oe) | References |
|---|---|---|---|---|---|---|---|
| | $\|\Delta S\|_{max}$ (J/K.kg) | T (K) | $\|\Delta T\|_{max}$ (K) | T (K) | | | |
| CFO (NFs) | 1.71 | 32 | 0.93 | 180 | 7.58 | 35000 | This work |
| Cobalt ferrite by co-precipitation (NPs) | 0.13 | 300 | - | - | - | 30000 | [50] |
| Cobalt ferrite sol-gel combustion method (NPs) | 0.23 | 213 | - | - | - | 13000 | [2] |
| $CoFe_2O_4/CoFe_2$ thermal reduction of $CoFe_2O_4$ (NPs) | 0.923 | 290 | - | - | 3.7 | 40000 | [1] |
| Zinc ferrite ball milling (NPs) | 0.30 | 130 | - | - | - | 30000 | [48] |

For the validation of the $CoFe_2O_4$ NFs as the magnetic refrigerant the most important parameter called the relative cooling power (RCP) is estimated and the cooling efficiency is given by

$$\mathrm{RCP} = |\Delta S_m(T,H)| \times \delta T_{FWHM} \quad \text{Eq(3) [7].}$$

where $\delta T_{FWHM}$ is the full width at half maximum of the magnetic entropy curve. RCP is a measure of the amount of heat transferred by a refrigerant per ideal cycle [3]. The best refrigeration cycles take place when the degree of RCP has a maximum value. Fig.10 shows the estimated relative cooling power changes to the external magnetic field. According to Fig. 10, the RCP value increases from 3.32 to 7.58 J/kg when the field increases from 10 to 35 kOe, with the large RCP value obtained only at 35 kOe.

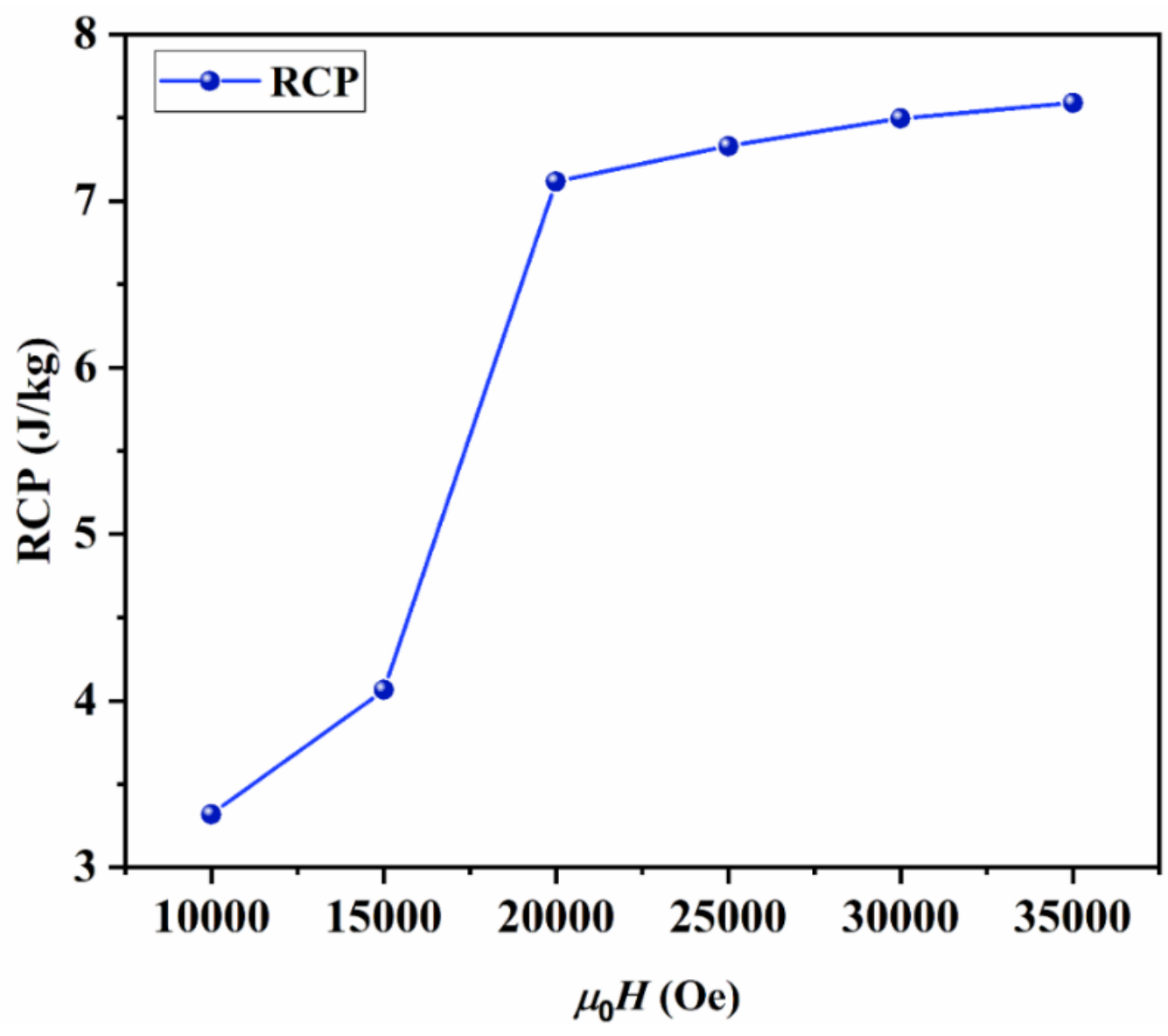

**Fig.10.** Relative cooling power of $CoFe_2O_4$ NFs.

## Conclusion

$CoFe_2O_4$ nanofibers have been successfully elaborated by the electrospinning technique. The MET and SEM show a continuous nanofibers morphology with an average grain size of about 210 nm. According to XRD, Raman, and SEAD, the CFO NFs crystallize in a pure cubic close-packed (c.c.p.) spinel structure with a space group of Fd $\overline{3}$ m. The ferromagnetic behavior of CFO is confirmed by M-H hysteresis loops. At ambient temperature, the magnetic field of 30 kOe resulted in $M_s$ and $H_c$ values of 75.87 emu/g and 723 Oe. The magnetocaloric effect on the NFs was estimated from the set of isothermal magnetization curves using Maxwell's relation for the first time. A high entropy change of $|\Delta S|$ =1.7 J/kg.K was observed around 32 K. The adiabatic temperature change $\Delta T_{max}$ = 0.93 K at T= 180 K, with the large relative cooling power of 7.58 J/kg at the field of 35 kOe. Overall, the controlled synthesis technique for one-dimensional CFO NFs significantly enhanced magnetic properties and magnetocaloric effect at lower temperatures. Such analysis would enhance the potential applications of these materials in the realms of nano electromagnetics and cooling devices.

### CRediT author statement

**Salma ElMouloua** : Investigation, Methodology, Data Curation, Writing - Original Draft, Validation, **Youness Hadouch** : Investigation, Methodology, Data Curation, Writing - Original Draft, Validation, **Salma Ayadh** : Visualization, Validation, **Salma Touili** : Review & editing, Validation, **Daoud Mezzane** : Visualization, Methodology, Writing – review & editing, Validation, Supervision, **M'barek Amjoud** : Visualization, Methodology, Writing – review & editing, Validation, Supervision, **Said Ben**

**Moumen** : Review & editing, Validation, **Alimoussa Abdelhadi** : Data processing, Visualization, Writing – review & editing, Validation, **Abdelilah Lahmar** : Formal analysis, Data processing, Visualization, Writing – review & editing, Validation, **Zvonko Jaglicic** : Formal analysis, Data processing, Visualization, Writing – review & editing, Validation, **Zdravko Kutnjak** : Visualization, Writing – review & editing, Validation, **Mimoun El Marssi** : Visualization, Writing – review & editing, Validation.

**Declaration of Competing Interest**

The authors declare that they have no known competing financial interests or personal relationships that could have appeared to influence the work reported in this paper.

**Data availability**

No data was used for the research described in the article

**Funding**

This work was supported by HORIZON-MSCA-2022-SE H-GREEN (No. 101130520), MSCA-2020-RISE-MELON (No. 872631).

**Acknowledgments**

The authors gratefully acknowledge the generous financial support of HORIZON-MSCA-2022-SE H-GREEN (No. 101130520), MSCA-2020-RISE-MELON (No. 872631).